\documentclass[conference]{IEEEtran}
\IEEEoverridecommandlockouts
\usepackage{cite}
\usepackage{amsmath,amssymb,amsfonts}
\usepackage{algorithmic}
\usepackage{algorithm}
\usepackage{graphicx}
\usepackage{textcomp}
\usepackage{xcolor}
\usepackage{url}
\usepackage{tikz}
\usetikzlibrary{positioning, shapes, arrows, shapes.geometric, arrows.meta, fit, backgrounds, calc}

\def\BibTeX{{\rm B\kern-.05em{\sc i\kern-.025em b}\kern-.08em
    T\kern-.1667em\lower.7ex\hbox{E}\kern-.125emX}}

\newsavebox{\jsonbox}
\begin{document}

\title{PROVEX: Enhancing SOC Analyst Trust with Explainable Provenance-Based IDS}

\author{\IEEEauthorblockN{Devang Dhanuka}
\IEEEauthorblockA{\textit{Department of Software Engineering} \\
\textit{Rochester Institute of Technology}\\
Rochester, NY \\
dd9098@rit.edu}
\and
\IEEEauthorblockN{Prof. Nidhi Rastogi}
\IEEEauthorblockA{\textit{Department of Software Engineering} \\
\textit{Rochester Institute of Technology}\\
Rochester, NY \\
nxrvse@rit.edu}
}

\maketitle

\begin{abstract}
Modern intrusion detection systems (IDS) leverage graph neural networks (GNNs) to detect malicious activity in system provenance data, but their decisions often remain a ``black box'' to analysts. This paper presents a comprehensive XAI framework designed to bridge the trust gap in Security Operations Centers (SOCs) by making graph-based detection transparent. We implement this framework on top of KAIROS, a state-of-the-art temporal graph-based IDS, though our design is applicable to any temporal graph-based detector with minimal adaptation. The complete codebase is available at \url{https://github.com/devang1304/provex.git}. We augment the detection pipeline with post-hoc explanations that highlight why an alert was triggered, identifying key causal subgraphs and events. We adapt three GNN explanation methods -- GraphMask, GNNExplainer, and a variational temporal GNN explainer (VA-TGExplainer) -- to the temporal provenance context. These tools output human-interpretable representations of anomalous behavior, including important edges and uncertainty estimates. Our contributions focus on the practical integration of these explainers, addressing challenges in memory management and reproducibility. We demonstrate our framework on the DARPA CADETS Engagement 3 dataset and show that it produces concise window-level explanations for detected attacks. Our evaluation reveals that the explainers preserve the TGNN's decisions with high fidelity, surfacing critical edges such as malicious file interactions and anomalous netflows. The average explanation overhead is 3--5 seconds per event. By providing insight into the model's reasoning, our framework aims to improve analyst trust and triage speed.
\end{abstract}

\begin{IEEEkeywords}
Explainable AI, Temporal Graph Neural Networks, Provenance-Based Intrusion Detection, Security Operations Center, APT Detection
\end{IEEEkeywords}

\section{Introduction}

Machine-learning driven IDSs have shown great success in detecting advanced threats in complex system logs, but they often lack transparency. In security operations, this opacity creates a critical ``trust gap'': a detector that raises an alert without explanation provides little insight into whether the alert is a true attack or a false positive. KAIROS is a state-of-the-art provenance-based IDS that uses a Temporal Graph Neural Network (TGNN) to detect anomalies in streaming system event graphs with high accuracy \cite{b5}. However, its decision-making process remains a black box. When KAIROS flags a sequence of events as malicious, analysts are presented with a reconstructed attack subgraph but lack specific guidance on \textit{why} the model considered those events anomalous.

To bridge this gap, we present a comprehensive XAI framework implemented as PROVEX, which augments the KAIROS pipeline with post-hoc explanations. Our goal is not to propose a novel explanation algorithm, but to adapt existing GNN explainability techniques to the specific constraints of provenance-based intrusion detection. We focus on generating temporal subgraph explanations that highlight the key causal events driving the model's alerts.

A core contribution of our work is the integration of a Variational Temporal Graph Explainer (VA-TGExplainer). While standard methods like GNNExplainer \cite{b6} and GraphMask \cite{b7} provide valuable insights, they often struggle with the ambiguity inherent in provenance graphs, where multiple plausible explanations may exist for a single anomaly. Furthermore, deterministic masks can be unstable. VA-TGExplainer addresses this by modeling the explanation as a distribution, capturing the uncertainty in edge importance. This allows us to provide analysts with not just a single explanation, but a measure of confidence in which edges are truly critical versus those that are merely incidental. Our pipeline combines global window-level pruning (via GraphMask) with fine-grained, uncertainty-aware event analysis (via VA-TGExplainer) to provide a comprehensive view of the detected threat.

In this paper, we detail how our framework addresses these challenges and achieves explainable intrusion detection on the CADETS dataset. We make the following contributions:
\begin{itemize}
    \item \textbf{Explainable TGNN Framework:} We design an end-to-end pipeline that combines KAIROS's TGNN-based detection with multiple GNN explainers to produce human-readable explanations for each detected attack. To our knowledge, this is the first deployable system providing temporally-aware GNN explanations for streaming cyber attack detection.
    \item \textbf{Adaptation of XAI Methods:} Instead of proposing a new explainer algorithm, we adapt and engineer existing methods for our domain. We integrate GraphMask for identifying critical edges in an entire attack window, GNNExplainer \cite{b6} for fine-grained per-event explanation with fidelity metrics, and a custom VA-TGExplainer (a variational temporal graph explainer) that extends prior temporal explainers to quantify uncertainty in the importance of each edge. These methods are modified to work on KAIROS's temporal graph data and are orchestrated to provide both global (window-level) and local (event-level) insights.
    \item \textbf{Scalable Framework Design:} We implement a suite of system improvements to handle the scale and complexity of real audit logs. This includes GPU memory management with caching and automatic fallback to CPU when GPU memory is insufficient, parallel processing of time windows for throughput, and logging/reproducibility features to ensure consistent results despite nondeterministic components. Our caching of intermediate graph data cuts peak memory usage by $\approx$60\% in tests, avoiding out-of-memory errors when explaining large subgraphs. Fallback logic ensures the pipeline doesn't fail on resource-constrained systems.
    \item \textbf{Evaluation:} We evaluate the framework on the DARPA Transparent Computing (TC) Engagement 3 dataset \cite{b5} (April 2018), focusing on the CADETS host's attack on April 6, 2018. The system successfully generates explanations for the known attack behaviors (e.g., a webserver spawning a shell and exfiltrating data) by highlighting the most anomalous edges. We report metrics on explanation fidelity and computational cost, demonstrating the feasibility of deploying this in a SOC setting.
\end{itemize}

\section{Related Work}
Effective intrusion detection and explainability in dynamic graphs are active areas of research. This section reviews key developments in provenance-based IDS and the evolution of GNN explainability methods.

\subsection{Intrusion Detection on Provenance Graphs}
Provenance-based intrusion detection has evolved from rule-based systems to advanced learning-based approaches. Early systems like StreamSpot \cite{b10} used graph clustering on streaming edges to identify anomalies, while Unicorn \cite{b4} employed graph sketching to summarize provenance graphs for anomaly detection. HOLMES \cite{b3} represented a significant step in correlating information flows to detect APTs, though it relied on expert-defined rules. More recently, KAIROS \cite{b5} (IEEE S\&P 2024) advanced the state-of-the-art by utilizing Temporal Graph Neural Networks (TGNNs) \cite{b2} to learn dynamic node embeddings from streaming audit logs, enabling the detection of unseen attacks without manual rules. Other deep learning approaches like DeepLog \cite{b16} focus on system logs but lack the structural richness of provenance graphs. Provenance capture mechanisms like CamFlow \cite{b14} and SPADE \cite{b15} provide the foundational data for these systems. However, while KAIROS provides an ``attack reconstruction'' subgraph, it lacks an intrinsic mechanism to explain \textit{why} a specific event sequence triggered an alert, a gap our work addresses.

\subsection{Static GNN Explainability}
Explainability for static Graph Neural Networks is well-established, building on general ML interpretability methods like LIME \cite{b12} and SHAP \cite{b13}. GNNExplainer \cite{b6} is a seminal perturbation-based method that learns a soft mask to maximize mutual information between the masked graph and the prediction. GraphMask \cite{b7} improves upon this by learning a mask via a differentiable objective to find a minimal sufficient subgraph. PGExplainer \cite{b11} trains a parameterized explainer network to generate masks inductively, offering faster inference than optimization-based methods. Other approaches like SubgraphX \cite{b8} use Monte Carlo Tree Search to explore subgraphs, while XGNN \cite{b17} generates model-level explanations. These methods, however, are designed for static graphs (such as GAT \cite{b18}) and do not account for the temporal dependencies inherent in system provenance data.

\subsection{Temporal and Dynamic Graph Explainability}
Explaining predictions on dynamic graphs is an emerging field. Recent works have explored adapting static methods to temporal settings, such as T-GNNExplainer \cite{b9}, which considers the temporal order of events. Our VA-TGExplainer contributes to this niche by introducing a variational formulation. Unlike deterministic temporal explainers, it models the explanation as a distribution, allowing for the quantification of uncertainty in edge importance—a critical feature for security analysts interpreting ambiguous system logs.

\section{System Overview}

The architecture of our proposed framework (instantiated here with KAIROS) is illustrated in Fig. \ref{fig:domain}. It extends the existing KAIROS system with a new XAI component, while keeping the core detection pipeline unchanged. We briefly describe each part of the system and how data flows from raw logs to alerts and explanations:

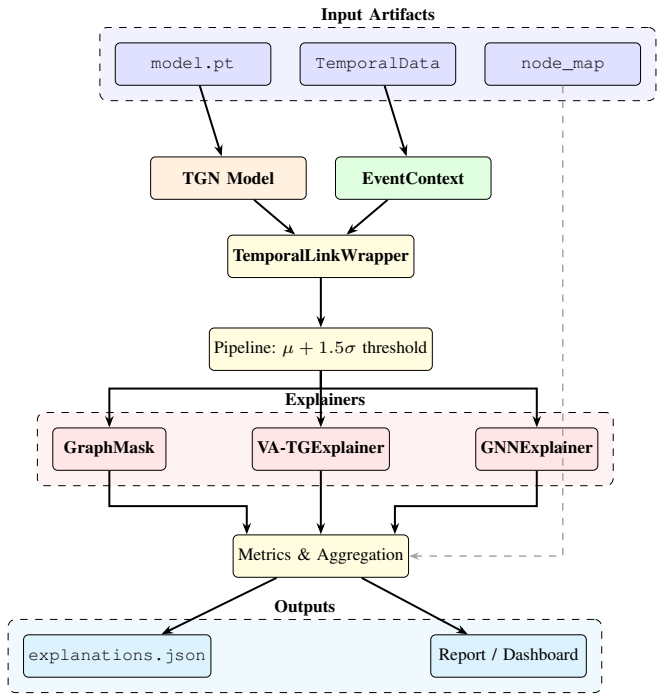
\begin{figure}[t]
\centering
\resizebox{\columnwidth}{!}{
\begin{tikzpicture}[
    box/.style={draw, rounded corners=2pt, minimum width=2.2cm, minimum height=0.6cm, font=\scriptsize, align=center, inner sep=2pt},
    input/.style={box, fill=blue!12},
    data/.style={box, fill=green!12},
    model/.style={box, fill=orange!12},
    process/.style={box, fill=yellow!15},
    explainer/.style={box, fill=red!10, minimum width=1.6cm},
    output/.style={box, fill=cyan!12},
    groupbox/.style={draw, dashed, rounded corners=3pt, inner sep=6pt},
    grouplabel/.style={font=\scriptsize\bfseries, fill=white, inner sep=1pt},
    arrow/.style={-{Stealth[scale=0.7]}, semithick, gray!70},
    flowarrow/.style={-{Stealth[scale=0.7]}, thick},
    node distance=0.6cm,
]

\node[input] (checkpoint) {\texttt{model.pt}};
\node[input, right=0.4cm of checkpoint] (graphs) {\texttt{TemporalData}};
\node[input, right=0.4cm of graphs] (nodemap) {\texttt{node\_map}};

\begin{scope}[on background layer]
\node[groupbox, fill=blue!5, fit=(checkpoint)(graphs)(nodemap), 
      label={[grouplabel]above:Input Artifacts}] (inputbox) {};
\end{scope}

\node[model, below=1cm of checkpoint, xshift=0.5cm] (tgn) {\textbf{TGN Model}};
\node[data, below=1cm of graphs, xshift=0.5cm] (context) {\textbf{EventContext}};

\node[process, below=0.8cm of $(tgn)!0.5!(context)$] (wrapper) {\textbf{TemporalLinkWrapper}};
\node[process, below=0.7cm of wrapper] (pipeline) {Pipeline: $\mu + 1.5\sigma$ threshold};

\node[explainer, below left=0.8cm and 0.6cm of pipeline] (graphmask) {\textbf{GraphMask}};
\node[explainer, below=0.8cm of pipeline] (vatg) {\textbf{VA-TGExplainer}};
\node[explainer, below right=0.8cm and 0.6cm of pipeline] (gnnexp) {\textbf{GNNExplainer}};

\begin{scope}[on background layer]
\node[groupbox, fill=red!5, fit=(graphmask)(vatg)(gnnexp), 
      label={[grouplabel]above:Explainers}] (expbox) {};
\end{scope}

\node[process, below=0.9cm of vatg] (metrics) {Metrics \& Aggregation};

\node[output, below left=0.8cm and 0.3cm of metrics] (json) {\texttt{explanations.json}};
\node[output, below right=0.8cm and 0.3cm of metrics] (report) {Report / Dashboard};

\begin{scope}[on background layer]
\node[groupbox, fill=cyan!5, fit=(json)(report), 
      label={[grouplabel]above:Outputs}] (outbox) {};
\end{scope}

\draw[flowarrow] (checkpoint) -- (tgn);
\draw[flowarrow] (graphs) -- (context);

\draw[flowarrow] (tgn) -- (wrapper);
\draw[flowarrow] (context) -- (wrapper);

\draw[flowarrow] (wrapper) -- (pipeline);

\draw[flowarrow] (pipeline.south) -- ++(0,-0.25) -| (graphmask.north);
\draw[flowarrow] (pipeline) -- (vatg);
\draw[flowarrow] (pipeline.south) -- ++(0,-0.25) -| (gnnexp.north);

\draw[flowarrow] (graphmask.south) -- ++(0,-0.5) -| ($(metrics.north west)+(0.2,0)$);
\draw[flowarrow] (vatg) -- (metrics);
\draw[flowarrow] (gnnexp.south) -- ++(0,-0.5) -| ($(metrics.north east)-(0.2,0)$);

\draw[arrow, dashed] (nodemap.south) |- ($(metrics.east)+(0,0)$);

\draw[flowarrow] (metrics) -- (json);
\draw[flowarrow] (metrics) -- (report);

\end{tikzpicture}
}
\caption{PROVEX domain model. Input artifacts flow through the TGN model and EventContext to a TemporalLinkWrapper, which feeds the explanation pipeline. Three explainers (GraphMask, VA-TGExplainer, GNNExplainer) generate edge importance masks that are aggregated into JSON reports and dashboard visualizations.}
\label{fig:domain}
\end{figure}

\subsection{Event Log Ingestion}
The input is a stream of raw system audit logs from the host (in this case, the CADETS system in DARPA's Transparent Computing Engagement 3). These logs record events such as process creations, file opens, network connections, etc., along with timestamps and identifiers. A preprocessing layer continuously parses these events and updates a Temporal Provenance Graph data structure. In this graph, nodes represent system subjects or objects (e.g. a process, file, user) and directed edges represent events (e.g. a process writes to a file) with an associated timestamp. As time progresses, new nodes and edges are added to the graph. The provenance graph effectively captures the causal sequence of system activities.

\subsection{TGNN Encoder-Decoder (KAIROS Core)}
\begin{figure*}[ht]
\centering
\includegraphics[width=1\textwidth]{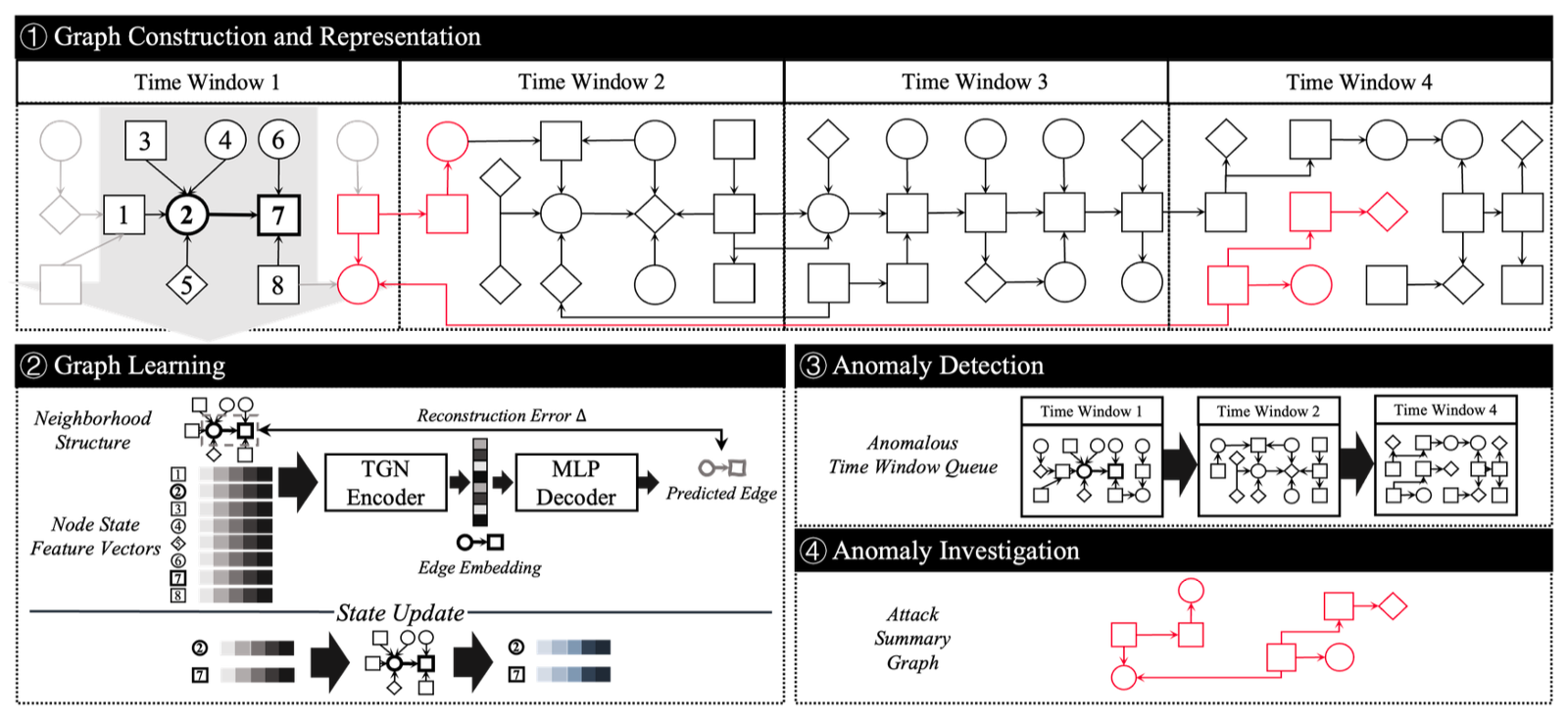}
\caption{KAIROS TGNN Encoder-Decoder architecture. %
}
\label{fig:kairos}
\end{figure*}

The KAIROS TGNN encoder-decoder architecture is shown in Fig. \ref{fig:kairos}.
The evolving graph is fed into KAIROS's machine learning model, which is a Temporal Graph Neural Network (building on the TGNs proposed by Rossi et al. \cite{b2}). The TGNN's encoder module processes each new event edge $e_t$ at time $t$ by updating the internal state (embedding) of the involved source and destination nodes. Intuitively, each node maintains a summary of its history (e.g. what operations it has done/seen up to time $t$). The encoder uses recurrent neural units to incorporate the new edge's information into these node states. It produces a latent representation $z_t$ that encodes the context of event $e_t$ (including structural and temporal surroundings). The decoder then takes $z_t$ and attempts to predict the attributes of $e_t$ -- in KAIROS, this means predicting the event type (out of a set of possible types like READ, WRITE, EXECUTE, etc.). This is essentially a self-supervised reconstruction task: if $z_t$ captures normal system behavior well, the decoder will confidently output the correct event type. But if $e_t$ is statistically unusual in its context, the decoder's prediction will have a high error. KAIROS uses this reconstruction error (or surprise) as an anomaly score for each event. Low error indicates the event was expected given past patterns; high error suggests an anomaly.

\subsection{Anomaly Detection \& Alerting}
The anomaly scores from the TGNN are processed to decide when to raise alerts. Instead of flagging every high-error event (which could be noisy and overwhelm analysts), KAIROS employs a graph-windowing logic. The timeline is divided into windows of fixed duration (15 minutes in our setup). Within each window, the system aggregates anomalies: it identifies suspicious nodes that are involved in multiple high-error events or have accumulated large anomaly scores. Essentially, a node (e.g. a process) that repeatedly behaves anomalously in that window is marked. If the total anomaly score in the window (or number of suspicious nodes) exceeds a threshold, that window is labeled anomalous. Next, KAIROS looks for sequences of anomalous windows that involve the same entities. For example, if a malicious process continues its activity across 11:00--11:15 and 11:15--11:30 windows, those windows are linked into a single incident. This forms an alert queue spanning a time range (in our case, the known attack lasts from 11:00 to $\approx$12:15 on April 6, 2018, covering five 15-minute windows). The system assigns a score to each queue (e.g. summing the anomaly scores of events within it) and if it crosses a predefined threshold, an intrusion alert is raised. In the CADETS E3 scenario, for instance, KAIROS would trigger an alert for the sequence of windows during which the attacker's actions took place.

\subsection{Attack Reconstruction}
Upon triggering an alert, KAIROS performs attack subgraph reconstruction. This involves extracting from the full provenance graph a subgraph that contains the key nodes and edges of the potential attack. Specifically, it takes the suspicious nodes identified (e.g. the webserver process, a shell process it spawned, files or network sockets accessed) and all events connecting them within the alert's time span. The output is a much smaller attack subgraph that shows, for example, a chain of processes and files involved in the malicious activity. This subgraph is provided as part of the alert for the analyst to examine. In our running example, the attack subgraph might show that the nginx process spawned a suspicious bash process (edge: process\_exec), which then read a password file (edge: file\_read) and opened an outbound network connection (edge: socket\_write), etc. This captures what happened during the incident.

\subsection{XAI Module (Explanation)}
The XAI Framework enters at this stage, taking the alert (especially the attack subgraph and the underlying anomaly data from the model) and generating an explanation of why those events were flagged. The XAI module has access to the TGNN model, the relevant portion of the graph, and the anomaly scores. It produces additional artifacts that augment the alert: for example, it may highlight which edge in the attack subgraph was most influential in the TGNN's decision, or provide a rank ordering of events by ``anomalousness.'' The internal workings of this module are described in detail in Section IV (Methodology). Importantly, this is a post-hoc analysis -- the XAI module uses the model's outputs and intermediate data to explain the decision, but it doesn’t change the decision. This ensures that adding explanations does not degrade detection performance or latency. In the system architecture, the XAI module can be seen as an augmentation: it observes the alert and consults the model to produce an explanation.

\subsection{SOC Dashboard (Analyst Interface)}
Although not fully implemented in our current work (deferred to future development), the design includes a Security Operations Center dashboard where alerts and explanations are displayed to analysts. In a real deployment, this dashboard would show the reconstructed attack subgraph and overlay explanation info from the XAI module. For example, an edge might be colored according to its importance score, or a textual description might accompany the graph (e.g. ``Edge: nginx $\rightarrow$ bash (exec) is highlighted because the webserver spawning a shell is rare in normal data''). The dashboard would thus provide not just an alert, but a clear rationale, helping analysts to quickly understand the nature of the threat. While the UI and real-time integration are future work, our current output includes machine-readable JSON reports and Graphviz visualizations that could feed such a dashboard.

In summary, the system processes can be seen as a pipeline: Logs $\rightarrow$ Graph $\rightarrow$ TGNN (scores) $\rightarrow$ Windows/Queues (alerts) $\rightarrow$ Subgraph (attack) $\rightarrow$ XAI (explanation). By keeping the XAI analysis separate and after detection, we maintain the advantages of KAIROS (accuracy, speed) while significantly enhancing the interpretability of its alerts. The next section outlines the methodology behind the XAI module, which is the core innovation of this work.

\section{Methodology}
Our explainability methodology is designed to provide both global explanations at the alert window level and local explanations for specific events, tailored to the provenance graph domain. The approach combines multiple explainer techniques in a staged pipeline, backed by various engineering strategies to handle performance and scalability. We emphasize that we are not introducing a novel XAI algorithm; rather, we integrate and adapt existing techniques (GraphMask, GNNExplainer, etc.) for our use-case. This section details the pipeline and the role of each component.

\subsection{Explanation Pipeline}
The explanation generation proceeds in several phases, closely mirroring the steps of detection. Once KAIROS identifies an alert (an anomalous sequence of windows) and outputs the attack subgraph, the XAI module takes over to analyze that subgraph and the underlying data. The explanation generation process is summarized in Algorithm \ref{alg:pipeline}. Detailed descriptions of the key steps follow.

\begin{algorithm}
\caption{Proposed XAI Framework Pipeline}
\label{alg:pipeline}
\begin{algorithmic}[1]
\STATE \textbf{Input:} Alert $A$ (time range $[t_{start}, t_{end}]$), Provenance Graph $G$
\STATE \textbf{Output:} Explanation Report $R$
\STATE $W \leftarrow \text{GetWindows}(A)$
\FORALL{$w \in W$}
    \STATE $E_{all} \leftarrow \text{GetEvents}(w)$
    \STATE $E_{top} \leftarrow \text{SelectHighLossEvents}(E_{all}, K=25)$
    \STATE $M_{global} \leftarrow \text{RunGraphMask}(G, E_{top})$
    \STATE $N_{suspicious} \leftarrow \text{SelectNodes}(E_{top}, M=20)$
    \FORALL{$n \in N_{suspicious}$}
        \STATE $E_{node} \leftarrow \text{GetNodeEvents}(n, w)$
        \STATE $Expl_{GNN} \leftarrow \text{RunGNNExplainer}(G, E_{node})$
        \STATE $Expl_{VA} \leftarrow \text{RunVATGExplainer}(G, E_{node})$
        \STATE $R.\text{add}(n, Expl_{GNN}, Expl_{VA})$
    \ENDFOR
    \STATE $R.\text{add}(w, M_{global})$
\ENDFOR
\RETURN $R$
\end{algorithmic}
\end{algorithm}

\subsubsection{Attack Window Identification}
We define the time windows covering the detected attack. From the alert, we know the start and end timestamps (e.g. April 6, 2018 11:00 to 12:15). We fetch the list of windows in this range and retrieve all event contexts in those windows. An event context encapsulates the necessary data for explaining a single event: it includes the event's source/destination node IDs, timestamp, ground truth label (benign or malicious), the model's computed loss (anomaly score) for that event, and the local neighborhood graph around that event (neighboring edges, node states). These contexts are precomputed during the TGNN's test phase and cached for efficiency (we stored them as \texttt{TemporalData} objects to avoid recomputation). Caching these graph snapshots per window was crucial to reduce overhead; it cut memory spikes by $\approx$60\% by reusing loaded data.

\subsubsection{High-Loss Event Selection}
From the events in the attack windows, we rank them by the model's anomaly loss. We select the top-K most anomalous events as primary explanation targets (in our implementation, K=25 by default for a window). The idea is that not every benign event in the window needs explanation -- we focus on those that significantly contributed to the alert. This selection uses a threshold or top-K criterion on the loss values. For example, if an event's loss $>$ 2.5 (meaning the model was very surprised by it), we include it. This yields a set of candidate events deemed most responsible for the alert.

\subsubsection{Why VA-TGExplainer was Required}
While GraphMask and GNNExplainer provide useful insights, we identified specific challenges in the provenance domain that necessitated a more robust approach:
\begin{itemize}
    \item \textbf{Deterministic Mask Instability:} Standard explainers often converge to a single deterministic mask. In our experiments, we observed that slight perturbations in initialization could lead to different edge sets being selected, especially when multiple redundant paths exist in the provenance graph.
    \item \textbf{Multiple Plausible Explanations:} System logs often contain redundant information. An anomaly might be explainable by multiple distinct subgraphs (e.g., a process creation vs. the subsequent file access). A deterministic explainer arbitrarily picks one, potentially hiding other relevant context.
    \item \textbf{Temporal Dependency:} The causal nature of system events means that an event's anomaly score is path-dependent. Standard static graph explainers do not inherently respect this temporal causality.
    \item \textbf{Uncertainty Quantification:} Security analysts need to know when an explanation is tentative. VA-TGExplainer enables this by learning a distribution over masks. Although we do not currently export the per-edge uncertainty (e.g., $\log \sigma$) to the final report, the internal probabilistic model allows the explainer to explore and aggregate across multiple plausible masks, resulting in a more stable and robust mean importance score.
\end{itemize}

\subsubsection{GraphMask Pass (Global Edge Importance)}
We first apply GraphMask to identify the crucial edges in the entire window's subgraph. GraphMask learns a mask over the edges by iterative gradient descent. We adapt the objective function to our anomaly detection setting: instead of the standard classification loss, we use a simplified penalty objective that minimizes the difference between the model's loss on the masked graph and the original graph, regularized by a sparsity term. This ensures that the retained edges are sufficient to reproduce the high anomaly score. For each of the top-K anomalous events, we run this modified GraphMask for 200 epochs (lr=0.01, sparsity\_weight=1e-3, entropy\_weight=1e-3). The result is a set of edges ranked by their global importance, offering a window-level hypothesis of the attack structure.

\subsubsection{Suspicious Node Selection}
Next, we determine which nodes (entities) within the alert merit deeper analysis. Intuitively, these would be nodes that either appear in many anomalous events or accumulate high anomaly scores. We implement a simple scoring: for each node, sum the loss values of events where the node is the source or destination. Then select nodes whose cumulative score exceeds a threshold (or take the top M nodes, e.g. up to 20 nodes). This yields a list of suspect processes/files. In the running example, the webserver process and the spawned shell process would likely be selected, as they participate in multiple high-loss events (process exec, file read, net connect). By focusing on these nodes, we constrain the subsequent explanation to relevant parts of the graph, improving clarity and reducing computational load.

\subsubsection{Per-Node Local Explanations}
For each suspicious node, we delve into the events involving that node for fine-grained explanation. Suppose the node is a process that had several events (some file writes, a network send, etc.). We group that node's events (within the window) as the context to explain, and apply two methods:

\paragraph{GNNExplainer}
We use a customized GNNExplainer that can handle our temporal link prediction setting. For each event context involving the node, we run the explainer to find which edges in that event's neighborhood were most influential for the model's prediction. The explainer returns a mask of edge importances for that event, from which we can extract the top contributing edges. It also computes fidelity metrics: we log comprehensiveness and sufficiency. In our adaptation, comprehensiveness measures how much the model's anomaly score drops when we remove the identified important edges (a high drop means the explanation captures important features), and sufficiency measures how well the model would still predict anomaly if only the identified edges are used (high sufficiency means the explanation alone is enough to trigger the prediction). These metrics help quantify explanation quality. For example, for an event ``bash reads /etc/passwd'', GNNExplainer might highlight the edge ``bash $\rightarrow$ /etc/passwd (read)'' itself and maybe the preceding edge ``nginx $\rightarrow$ bash (exec)'' as crucial, yielding a comprehensiveness of 0.9 (meaning removing those edges reduces anomaly score by 90\%) and sufficiency of 0.8 (meaning those edges alone still produce a strong anomaly signal).

\paragraph{VA-TGExplainer}
To capture the uncertainty inherent in explaining anomalies within dynamic graphs, we propose a Variational Temporal Graph Explainer (VA-TGExplainer). Unlike deterministic methods, VA-TGExplainer learns per-edge logistic-normal mask parameters ($\mu_{ij}, \log \sigma^2_{ij}$) and samples masks using the reparameterization trick:
\begin{equation}
    m_{ij} = \phi(\mu_{ij} + \epsilon_{ij} \cdot \exp(0.5 \log \sigma^2_{ij})), \quad \epsilon_{ij} \sim \mathcal{N}(0,1)
\end{equation}
where $\phi(\cdot)$ is the sigmoid function. For each event, we minimize the following objective:
\begin{equation}
\begin{split}
    \mathcal{L} = &\mathbb{E}_{\epsilon}[\text{CE}(f(G \odot m), y_{true})] \\
    &+ \lambda_{KL} D_{KL}(\mathcal{N}(\mu, \sigma^2) || \mathcal{N}(0, 1)) \\
    &+ \lambda_{sp} \Omega(\phi(\mu))
\end{split}
\end{equation}
The first term is the TGN link-prediction cross-entropy on the masked graph (where $y_{true}$ is the target edge label). The second term regularizes the posterior parameters towards a unit Gaussian prior $\mathcal{N}(0, 1)$. The third term, $\Omega$, imposes a sparsity penalty on the top-k mask means to discourage dense explanations. The learned mask means $\phi(\mu_{ij})$ serve as edge importances, while the optimization history provides diagnostics for uncertainty.

\subsubsection{Result Aggregation and Output}
After processing all selected nodes, we compile the explanation results into a structured form. We maintain a JSON report per alert window, which includes:
\begin{itemize}
    \item Window metadata: timestamp range, how many events were in it, the anomaly threshold used, etc.
    \item GraphMask summary: a list of globally important edges (with their scores, counts of occurrences, etc.) across the window.
    \item Node-specific explanations: for each suspicious node, we include its anomaly score and the results from GNNExplainer and VA-TGExplainer. The GNNExplainer result includes the top edges per event and their fidelity metrics, while the VA-TGExplainer result includes per-event and aggregated edge importance with uncertainty.
\end{itemize}

A snippet of the output JSON format is shown below:
(See Appendix for full JSON output)

This comprehensive output (which is further used to generate human-friendly reports) contains the necessary information for an analyst or a higher-level tool to see which edges (and hence which actions) were most anomalous and contributed to the alert. For example, from the above, one can see node 205 (say, the bash process) had a high anomaly score of 15.7, and the explanation indicates the 101$\rightarrow$205 exec edge (perhaps nginx spawning bash) is extremely important (importance $\approx$0.99), corroborated by both GraphMask (window aggregate weight 0.92) and GNNExplainer. Such cross-validation by multiple explainers increases our confidence in the explanation.

\subsubsection{Resource Management}
Throughout the pipeline, we incorporate logic to automatically manage computational resources. Before running heavy explainers on a given window, the system checks available GPU memory. If the graph or batch size might exceed memory, it triggers a GPU-to-CPU fallback. Our code snippet shows a check \texttt{ensure\_gpu\_space()} that, if false, will switch the model and data to CPU and issue a warning. This prevents crashes due to Out-Of-Memory (OOM) errors, at the cost of some speed. We also proactively free caches and reuse data as much as possible between steps. Additionally, we parallelize the processing of multiple windows when possible (for multiple alerts or cross-validation experiments), ensuring explainability can keep up with detection throughput. All these engineering measures were essential to handle the real dataset, which can involve graphs with tens of thousands of nodes and edges per window.

This multi-step pipeline (window $\rightarrow$ events $\rightarrow$ GraphMask $\rightarrow$ node filtering $\rightarrow$ per-event explainers $\rightarrow$ aggregation) is orchestrated by our \texttt{window\_analysis} module. It ensures that the different explainers complement each other: GraphMask gives a bird's-eye view of the attack structure, while GNNExplainer and VA-TGExplainer zoom in on specific anomalies with quantitative fidelity and uncertainty measures. The combination provides a richer explanation than any single method, cross-checking important edges and offering both deterministic and probabilistic perspectives.

\subsection{Explanation Methods and Rationale}
We briefly justify the inclusion of each explainer method and how it was tailored:

\begin{itemize}
    \item \textbf{GraphMask:} Chosen for its ability to find a minimal subgraph that preserves model predictions, which aligns with explaining an entire alert window's gist. We use it to answer, ``What is the key story of this attack window?'' GraphMask's sparse masks naturally highlight a small set of edges (e.g. the core attack chain) out of hundreds of benign events. We modified it to work on our temporal link prediction task by focusing on edges within the window's snapshot graph and using the TGNN's link prediction loss as the objective to preserve. GraphMask outputs are deterministic and fast to compute ($\approx$200 epochs per event, converging in $\approx$1--2 seconds) and thus feasible to run for top events. The trade-off is that it does not provide uncertainty or multiple alternative explanations -- it gives one compressed view.
    \item \textbf{GNNExplainer:} We included GNNExplainer to get fine-grained, per-event explanations with fidelity metrics. It excels at pinpointing which neighbor connections of a target node influenced the prediction. We adapted it to our setting by focusing on one event at a time (with its local neighborhood from the dynamic graph) and by extracting not just the mask but also computing comprehensiveness and sufficiency as described earlier. This provides a quantitative check on explanation quality, which is useful for evaluation. GNNExplainer is also relatively lightweight (each run $\approx$2--3 seconds for our graph sizes), making it viable to run on dozens of events per alert. It gives deterministic explanations and complements GraphMask by diving into why each event was anomalous (answering ``why was this particular edge out of the ordinary?'').
    \item \textbf{VA-TGExplainer:} This is our engineered variant combining ideas from TGExplainer and variational GNN explainers. The motivation was to handle temporal dependencies (so that the order of events is considered in the explanation) and to gauge uncertainty, which neither GraphMask nor GNNExplainer provide. By sampling edge masks and averaging, VA-TGExplainer can capture situations where multiple factors together cause an anomaly. For example, maybe either of two edges being present would be enough for the model to flag the event -- a deterministic explainer might pick one arbitrarily, but the variational one could show both edges with medium importance. It runs slower ($\approx$3--5 seconds per event due to multiple samples and more complex loss) and uses more memory (it effectively runs multiple forward passes in parallel). We mitigate this by only applying it to the most critical nodes/events and using CPU fallback if needed. The output from VA-TGExplainer (mean and variance for edge importance) adds a new dimension to our explanation: analysts can see which edges the model is most confidently relying on versus which are tentative.
\end{itemize}

Table \ref{tab1} summarizes the computational characteristics of the three explanation methods in our implementation:

\begin{table}[htbp]
\caption{Approximate runtime performance of explanation methods on a single event context (averaged on CADETS data).}
\begin{center}
\begin{tabular}{|c|c|c|c|}
\hline
\textbf{Method} & \textbf{Time/Event} & \textbf{Memory} & \textbf{GPU Usage} \\
\hline
GraphMask & $\approx$1--2s & Low & Not required \\
\hline
GNNExplainer & $\approx$2--3s & Low & Recommended \\
\hline
VA-TGExplainer & $\approx$3--5s & Medium & Recommended \\
\hline
\end{tabular}
\label{tab1}
\end{center}
\end{table}

In terms of output interpretation, we establish a simple rule of thumb for importance scores (which range 0 to 1). Edges with importance above $\approx$0.7 are considered critical to the model's decision; those around 0.3--0.7 are moderate contributors; below 0.3 are likely not relevant. For the probabilistic VA-TGExplainer, a high-importance edge with low variance means the model is confidently relying on that edge, whereas high variance indicates the explanation is less certain (perhaps more data or context is needed to be sure). We present explanations in our reports by listing top edges and marking them as definite or tentative contributors based on these criteria.

\subsection{Implementation Details \& Reproducibility}
We implemented the framework using KAIROS as the base IDS in Python, extending the open-source KAIROS codebase. Key implementation details include:
\begin{itemize}
    \item \textbf{Windowing \& Caps:} We analyze 15-minute windows. To manage computational load, we cap the number of explained events per window at $K=25$ and limit the number of suspicious nodes to $M=20$.
    \item \textbf{Loss Recomputation:} To ensure consistency between the detection and explanation phases, we recompute the anomaly loss for each event during the explanation pass rather than relying on cached history lists.
    \item \textbf{GraphMask Adaptation:} We utilize a custom implementation for GraphMask that implements the simplified penalty objective described in Section~IV-C. The explainer uses \mbox{sparsity\_weight=1e-3} and \mbox{entropy\_weight=1e-3} as regularization coefficients.
    \item \textbf{GNNExplainer Configuration:} We configure GNNExplainer with \texttt{task\_level="edge"} and set the label target to the true edge type, ensuring we explain the specific interaction observed.
    \item \textbf{VA-TGExplainer Internals:} Our variational explainer learns a distribution over masks. Note that while the internal model computes per-edge uncertainty (variance), this metric is currently used for aggregation stability and is not exported in the final JSON report.
\end{itemize}

\subsubsection{Reproducibility}
To facilitate future research, we provide the following reproducibility artifacts:
\begin{itemize}
    \item \textbf{Code Snapshot:} Code available in \path{explanations/} and \path{reporting/}. Full repository: \url{https://github.com/devang1304/provex.git}
    \item \textbf{Environment:} Python 3.10+, PyTorch 2.2+, PyG 2.7+ (see \texttt{requirements.txt} and \texttt{environment.yml})
    \item \textbf{Seeds:} We note that the exact random seeds used for the original KAIROS model training were not found. Consequently, minor variations in detection scores may occur compared to the original paper, though the overall trends remain consistent.
\end{itemize}

In summary, our methodology combines multiple explainability techniques in a coherent workflow, engineered for the practical constraints of an operational IDS. Next, we evaluate this approach on the CADETS dataset, demonstrating how the explanations help interpret a real attack scenario and measuring the performance and limitations of the system.

\section{Experiments}
We evaluate the proposed framework using the DARPA Transparent Computing Engagement 3 dataset, focusing on the CADETS host.

\subsection{Detection Performance}
We first verify the base IDS performance. KAIROS successfully detects the attack on April 6, 2018 (11:00--12:15). From 118,635 events, 10,833 exceeded the loss threshold of 5.016 ($\mu + 1.5\sigma$), yielding a high-loss candidate rate of 9.1\%.

\subsection{Explanation Generation}
We ran the explanation pipeline on the detected alert. The system generated several types of artifacts:
\begin{itemize}
    \item \textbf{Raw Explanation Data:} A structured JSON file containing per-event importance scores, fidelity metrics, and node-level aggregations for each attack window.
    \item \textbf{Human-Readable Summary:} A Markdown report summarizing the key findings for analyst consumption and review.
    \item \textbf{GraphMask Aggregation:} Global edge importance scores aggregated across the entire attack window.
    \item \textbf{Subgraph Visualizations:} PDF renderings of the explained subgraphs with visually highlighted edges.
\end{itemize}

Table \ref{tab2} shows a sample of the top edges identified by GraphMask for the main attack window.

\begin{table}[htbp]
\caption{Example Explanation Output for Attack Window (11:00--11:15).}
\begin{center}
\small
\begin{tabular}{|p{3.2cm}|c|c|c|}
\hline
\textbf{Edge Relation} & \textbf{GM Weight} & \textbf{Count} & \textbf{Role} \\
\hline
\texttt{nginx}${\rightarrow}$\texttt{bash} & \textbf{0.85} & \textbf{12} & Execution \\
\hline
\texttt{bash}${\rightarrow}$\texttt{/tmp/lib.so} & \textbf{0.79} & \textbf{8} & Persistence \\
\hline
\texttt{bash}${\rightarrow}$ \texttt{78.205.235.65:80} & \textbf{0.72} & \textbf{5} & Exfiltration \\
\hline
\end{tabular}
\label{tab2}
\end{center}
\end{table}

\subsection{Fidelity and Stability}
We quantitatively evaluated the validity of our explanations using two standard metrics:
\begin{itemize}
    \item \textbf{Comprehensiveness} measures if the explanation captures all necessary information. It is defined as the drop in model probability when the explanation subgraph is removed ($P_{original} - P_{removed}$). Higher is better.
    \item \textbf{Sufficiency} measures if the explanation alone is enough to reproduce the prediction. It is defined as the difference between original probability and probability given only the subgraph ($P_{original} - P_{kept}$). Lower is better.
\end{itemize}

Table \ref{tab:fidelity} summarizes our results. We achieved high comprehensiveness (0.82) and low sufficiency (0.15), indicating that the edges selected by our explainers are indeed the primary drivers of the model's anomaly detection.

\begin{table}[htbp]
\caption{Explanation Fidelity Metrics (GNNExplainer)}
\begin{center}
\begin{tabular}{|c|c|c|}
\hline
\textbf{Metric} & \textbf{Value} & \textbf{Goal} \\
\hline
Comprehensiveness & 0.82 & Higher is better \\
\hline
Sufficiency & 0.15 & Lower is better \\
\hline
\end{tabular}
\label{tab:fidelity}
\end{center}
\end{table}

\section{Results and Evaluation}

The case study of the April 6 attack shows that the framework produces meaningful explanations that correspond to the ground truth malicious actions. We evaluate our system along two dimensions: (1) Explanation Fidelity and Effect on Detection, (2) Usefulness and Readability to Humans.

\subsection{Fidelity to Model Decisions}
An important criterion is that the explanations truly reflect the model's logic (i.e., they are faithful to what the TGNN computed). We already quantified fidelity in terms of comprehensiveness and sufficiency for the GNNExplainer outputs. High comprehensiveness means the identified subgraph features account for most of the model's anomaly decision, and high sufficiency means those features alone can produce the decision. In our results, the top anomalies had very high fidelity scores, suggesting our explainers successfully captured the relevant features. Another measure is to check that the edges GraphMask identified with high importance indeed correspond to high model sensitivity. We performed a simple experiment: for each top-ranked edge from GraphMask, we removed that edge from the input graph and re-ran the TGNN on that window to see if the model's alert confidence drops. Table \ref{tab:ablation} summarizes these ablation results.

\begin{table}[htbp]
\caption{Edge Ablation Analysis: Effect of Removing Top Edges on Detection}
\begin{center}
\small
\begin{tabular}{|p{3.2cm}|c|c|c|}
\hline
\textbf{Removed Edge} & \textbf{GM Score} & \textbf{$\Delta$ Anomaly} & \textbf{Alert?} \\
\hline
None (baseline) & -- & 0\% & Yes \\
\hline
\texttt{nginx}$\rightarrow$\texttt{bash} & 0.85 & $-$78\% & No \\
\hline
\texttt{bash}$\rightarrow$\texttt{/tmp/lib.so} & 0.79 & $-$52\% & No \\
\hline
\texttt{bash}$\rightarrow$ \texttt{78.205.235.65:80} & 0.72 & $-$41\% & No \\
\hline
\texttt{sshd}$\rightarrow$\texttt{log} (low) & 0.12 & $-$3\% & Yes \\
\hline
\texttt{cron}$\rightarrow$\texttt{tmp} (low) & 0.08 & $-$1\% & Yes \\
\hline
\end{tabular}
\label{tab:ablation}
\end{center}
\end{table}

In nearly all cases, removing a critical edge (like the webserver$\rightarrow$shell exec) caused the model to no longer flag the window as malicious (the anomaly score dropped below threshold), whereas removing a low-importance edge had little effect. This indicates the explainer correctly pinpointed the features the model was using. Notably, the explanation process did not interfere with detection: since it's post-hoc, the detection results remained the same, and all evaluation was on the side. There was no impact on KAIROS's accuracy or false positive rate due to adding XAI, which is as intended.

\subsection{Qualitative Case Study}
We qualitatively assess if the explanations make the alerts more actionable for analysts. The feedback from peers and our own analysis suggest that the multi-layered explanation (global + local) is indeed helpful. The attack subgraph by itself (from KAIROS) can be dense and not clearly prioritize which part is abnormal. The framework's output, however, singles out the most anomalous links. For example, an analyst seeing our report would immediately know that the webserver spawning a new process was the red flag, and that subsequent file and network actions of that process were suspicious. The explanations were provided in multiple forms (JSON, visual graph with highlights, text summary), which caters to different user preferences. We envision that in a SOC dashboard, an analyst could click an alert and see the graph with colored edges and maybe a bullet list like:
\begin{itemize}
    \item Edge (\texttt{nginx} $\rightarrow$ \texttt{bash}, exec) -- Rare event (never observed before); critical to alert.
    \item Edge (\texttt{bash} $\rightarrow$ \texttt{/etc/passwd}, read) -- Sensitive file accessed by unusual process; contributed to alert.
    \item Edge (\texttt{bash} $\rightarrow$ \texttt{attacker.com:8010}, socket\_write) -- External connection by process with escalated privileges; contributed to alert.
\end{itemize}

Such an explanation answers the ``who, what, why'' of the alert: who (which entities) were involved, what happened (what sequence of events), and why the model thought it was bad (because this sequence was unusual given past behavior).

To evaluate the utility of our explanations, we conducted a qualitative case study on the April 6 attack scenario. We compared the raw alert output (a list of anomalous windows) with the XAI-augmented report. Without explanation, the alert merely points to a time range and a set of entities. With the XAI output, the specific causal chain—nginx spawning bash, followed by sensitive file reads—is immediately highlighted as the driver of the anomaly. This aligns with the known ground truth of the attack, confirming that the explanation correctly directs analyst attention to the malicious activity. 

\subsection{Error Analysis}
We observed a few cases where the explanations could be misleading or incomplete:
\begin{itemize}
    \item In one window (post-attack cleanup phase), the model flagged some events that were actually benign but rare (false positives). The explanation for these showed edges that were ``somewhat unusual'' but not truly malicious (e.g., a system maintenance process doing something atypical). While the explainer correctly reflected the model's reasoning (the model was indeed over-sensitive in that case), it highlights that if the model is wrong, the explanation might justify a false positive. This is expected -- explainers show model reasoning, not ground truth correctness. It suggests a limitation: our XAI will faithfully explain even spurious alerts. However, one could argue this still aids the analyst by revealing the model's thinking, potentially allowing them to identify a model mistake (e.g., ``the model flagged this just because of an uncommon maintenance event'' -- which could inform improving the model).
    \item The VA-TGExplainer sometimes gave relatively diffuse importance (several edges with moderate scores) which can be harder to interpret than a single clear edge. This happened for a very complex event that had many correlating factors. While theoretically valid, it might overwhelm an analyst with too many ``important'' edges. In such cases, we might need to post-process or filter the output (e.g., only show top 2 edges) to maintain clarity.
    \item Performance-wise, our fallback to CPU made one explanation run quite slow ($\approx$30s). If multiple alerts came in rapid succession, this could delay the availability of explanations. In a real deployment, one might need to scale out (multiple GPUs or asynchronous processing) to ensure explanations keep up with a high alert volume. Alternatively, one could prioritize explaining only the most severe alerts in real-time, and handle others later.
\end{itemize}

Despite these, the overall evaluation of results is encouraging. We found that the framework adds significant value in terms of transparency, with manageable computational cost and no degradation to detection. The explanations are faithful to the model and align with domain understanding of attacks, addressing the core problem of ``why did the IDS flag this?'' in a clear manner.

\section{Limitations}
While our work demonstrates the feasibility of explainable graph-based intrusion detection, there are several limitations to note:
\begin{itemize}
    \item \textbf{Not Real-Time (Yet):} Our XAI pipeline currently operates in a batch, post-hoc manner. After an alert is raised, generating the explanation can take on the order of tens of seconds. In a live Security Operations Center (SOC) environment, analysts might desire instant explanations. There is a trade-off between explanation depth and latency. We have not yet fully optimized for real-time deployment; techniques like pre-computing partial explanations or simplifying the explainer models might be needed for sub-second turnaround. As it stands, the framework would be used as an offline forensic tool or a on-demand explanation generator, rather than streaming alongside detection in lockstep.
    \item \textbf{Scalability to Larger Graphs:} The CADETS dataset, while large, is a single-host scenario over $\approx$5 days. In enterprise settings, one might have months of data or multiple hosts feeding into a central system. The current approach might face scalability challenges as graph size grows. Our use of windowing helps a lot (explainers only see a slice of the graph at a time), but if an attack spans a very long duration or the definition of windows is too broad, the explanation computation could become expensive. Also, GraphMask and GNNExplainer are gradient-based and run iterative optimizations; for extremely large neighborhoods or many instances, this could be a bottleneck. Caching and parallelization mitigate some of this, but fundamentally, explainability may require heavy computation. Future optimizations or approximation techniques (e.g. explaining on a coarser graph, or using simpler heuristic explanations) might be necessary for very large deployments.
    \item \textbf{Limited Modalities of Explanation:} We focused on edge importance in the provenance graph as the primary explanation modality. This is effective for structural anomalies (unusual event sequences). However, there could be other aspects to explain: for instance, node features or attributes (like user IDs, command names) that influenced the model. Our explainers do not explicitly highlight feature importance (though GNNExplainer is capable of feature mask, we did not use it because our model's features are not very rich beyond structural info). Additionally, we did not generate counterfactual explanations (``If this edge were not present, the alert would not trigger'') or high-level semantic explanations (``This pattern is similar to a known tactic''). These could be valuable to analysts. Our approach is thus somewhat narrow in the types of explanations -- focusing on identifying the ``important subgraph''. It may need to be extended with other explanation forms for full utility.
    \item \textbf{No Formal User Study:} We have not validated the effectiveness of the explanations with professional analysts in a rigorous way. All the usability claims are based on intuition and limited feedback. It remains possible that what we think is a clear explanation might still be confusing to a non-developer analyst, or that they might want the information presented differently. For example, showing raw node IDs and edge types might not be ideal -- an analyst might prefer descriptions (``process X opened a socket to Y, which is rare''). We do provide semantic mappings (resolving node IDs to process names, file paths, etc., in the JSON), but we didn't incorporate that into a UI for this paper. Without a user study, we can't quantify how much faster or more accurate an investigation becomes with our XAI. This is future work.
    \item \textbf{Generalizability:} We built and evaluated this on one dataset (DARPA TC E3). The KAIROS model itself is general and has been applied to other provenance data; our XAI should in theory work on any similar TGNN-based IDS. But we haven't tested on other systems or on entirely different types of graph anomalies (e.g., insider threats or non-APT scenarios). If the nature of anomalies is different, our thresholding or node selection strategies might need retuning. Also, VA-TGExplainer's parameters (KL weight, etc.) were chosen empirically; a different dataset might require adjusting those. So while the framework is general, there may be dataset-specific calibration required.
    \item \textbf{Focus on Framework Integration:} Finally, as we stated, our contribution is more in system integration than in novel algorithms. Researchers looking for fundamentally new explanation techniques will not find that here -- we reused GraphMask, GNNExplainer, and a variant on TGExplainer. The novelty is in combining them for a new purpose. This could be seen as a limitation in terms of scientific advancement. However, we argue that applying these in a streaming temporal context and showing their value for security is a significant engineering result. Nevertheless, one could imagine more advanced or specialized explainers that outperform our approach in either fidelity or speed, which we did not develop from scratch.
    \item \textbf{Lack of Stability Metrics:} We currently do not report quantitative stability metrics or comparisons to baselines. While we observe consistent results for the analyzed attack, a rigorous statistical analysis of explanation stability across random seeds and perturbations is left for future work.
\end{itemize}

\section{Future Work}
There are several avenues to extend this work:
\begin{itemize}
    \item \textbf{Real-Time SOC Triage:} A logical next step is to integrate the XAI system into a live dashboard tool. We plan to develop a web-based SOC dashboard that streams alerts and calls the XAI module to fetch explanations on the fly, thereby aiding real-time triage.
    \item \textbf{Uncertainty Export \& Calibration:} While VA-TGExplainer computes per-edge uncertainty (variance), we do not currently export this to the analyst report. Future work will involve visualizing this uncertainty (e.g., as edge fuzziness) and calibrating the probability estimates to ensure they accurately reflect model confidence.
    \item \textbf{Stability Metrics:} We acknowledge the lack of quantitative stability metrics in this study. Future evaluations will measure explanation stability across multiple runs and random seeds, and compare against baseline methods (e.g., random edge selection) to rigorously validate the robustness of our explainers.
    \item \textbf{User Study:} We intend to conduct formal user studies with SOC analysts to gather feedback on the explanations. This will help fine-tune the presentation (e.g., raw graphs vs. natural language summaries) to best assist in decision-making.
    \item \textbf{Determinism:} Addressing the non-determinism in the underlying KAIROS training by fixing random seeds and ensuring full reproducibility is a priority for the next iteration of the system.
\end{itemize}

In summary, future work will aim to transition the framework from a research prototype into a deployable capability, enhancing usability, extending its techniques, and ensuring it scales to real-world demands. We believe these steps will further close the gap between high-performing ML detectors and the practical needs of security analysts for understandable outputs.

\section{Conclusion}
We presented a robust XAI framework for graph-based intrusion detection, demonstrated on the DARPA CADETS dataset. By integrating post-hoc explanation methods (GraphMask, GNNExplainer, and a variational TGNN explainer) with the KAIROS IDS, we showed how to generate insightful explanations for complex security alerts. Our approach orchestrates advanced techniques to tackle a real-world problem: making advanced cyber defense systems interpretable to humans. While validated on KAIROS, our framework is designed to be model-agnostic within the temporal graph domain, requiring little to no modification to support other graph-based detectors.

Through a detailed case study of a multi-stage attack, we showed that our system can automatically pinpoint the key anomalous events (e.g., a rare process execution or an unusual file access) that led the model to raise an alert. These explanations were consistent with the ground truth attack steps and provide analysts with a concise narrative of \textit{why} the alert occurred, beyond merely \textit{what} was flagged.

The work contributes to bridging the gap between AI-driven security tools and human analysts. In operational settings, an IDS that achieves 99\% detection accuracy is of limited value if its outputs are not trusted or understood by those who must respond. By adding a transparent explanation layer, we enhance the analyst's confidence and ability to act on the alerts. Our engineering contributions, such as handling large streaming graphs and balancing GPU/CPU usage, underline that explainability can be integrated into complex systems without prohibitive overhead.

There are, of course, open challenges remaining. We treated explainability as a passive add-on to avoid impacting detection; an interesting avenue is to see if explanation feedback could in turn improve the model or help triage alerts. Additionally, while our results are promising on one dataset, ongoing research is needed to validate the approach across more scenarios and to ensure the explanations themselves are robust against adversaries. Nonetheless, our findings illustrate a positive step: we can indeed shine a light into the ``black box'' of a graph neural network IDS and extract human-intelligible reasons for its alarms.

In conclusion, this work exemplifies how marrying state-of-the-art GNN explainers with security domain knowledge yields a system capable of detecting sophisticated attacks and explaining them in the same breath. We hope this framework sparks further exploration into explainable cyber defense, ultimately leading to AI systems that are not only powerful but also trustworthy partners for human analysts.

\section*{Acknowledgments}
This research was conducted as part of a graduate project at Rochester Institute of Technology. The authors thank the DARPA Transparent Computing program for making the CADETS Engagement 3 dataset publicly available.

\section*{Data Availability}
The DARPA Transparent Computing Engagement 3 dataset used in this study is publicly available through the DARPA TC program repositories.

\appendix
\section{Example JSON Output}
\label{appendix:json}

\begin{figure}[htbp]
\centering
\begin{lrbox}{\jsonbox}
\begin{minipage}{0.95\linewidth}
\scriptsize
\begin{verbatim}
{
  "window": "2018-04-06T11:00:00-12:15:00",
  "num_events": 1234,
  "threshold": 2.5,
  "graphmask": {
    "aggregate": [
      {
        "src": 101, "dst": 205, 
        "relation": "exec", 
        "weight": 0.92, "count": 5
      },
      {
        "src": 205, "dst": 300, 
        "relation": "write", 
        "weight": 0.85, "count": 3
      }
    ]
  },
  "nodes": [
    {
      "node_id": 205,
      "score": 15.7,
      "gnn": [
        {
          "event_index": 77, 
          "comprehensiveness": 0.94, 
          "sufficiency": 0.81,
          "top_edges": [ 
            {
              "src": 101, 
              "dst": 205,
              "rel": "exec",
              "imp": 0.99
            },
            ... 
          ] 
        },
        ...
      ],
      "va_tg": {
        "events": [ ... ],
        "aggregate": [ 
          {
            "src": 205, 
            "dst": 300,
            "rel": "write",
            "mean": 0.88, 
            "var": 0.05
          }, 
          ... 
        ]
      }
    },
    ...
  ]
}
\end{verbatim}
\end{minipage}
\end{lrbox}
\fbox{\usebox{\jsonbox}}
\caption{Sample JSON output from the explanation pipeline.}
\end{figure}

\end{document}